\documentclass[apjl]{emulateapj}

\usepackage{natbib}
\newcommand{\be}{\begin{equation}}
\newcommand{\ee}{\end{equation}}
\newcommand{\bea}{\begin{eqnarray}}
\newcommand{\eea}{\end{eqnarray}}

\begin{document}

\title{Optimal Softening for $N$-Body Halo Simulations}
\author{Hu Zhan} 
\shortauthors{Zhan}
\shorttitle{Optimal Softening}
\affil{Department of Physics, University of California, Davis, CA 
       95616}
\email{zhan@physics.ucdavis.edu}

\begin{abstract}
We propose to determine the optimal softening length 
in $N$-body halo simulations by minimizing the 
ensemble-average acceleration error at a small radius $r_0$. This
strategy ensures that the error never exceeds the optimal value 
beyond $r_0$. Furthermore, we derive semi-analytic formulae for 
calculating the acceleration 
error due to the discreteness of particles and softened gravity,
which are validated by direct $N$-body force calculations.
We estimate that current state-of-the-art halo simulations 
suffer $\gtrsim 6\%$ acceleration error at $1\%$ of the halo
virial radius. The error grows rapidly toward the center
and could contribute significantly to the uncertainties of  
inner halo properties.
\end{abstract}


\section{Introduction} \label{sec:intr}
$N$-body simulations use discrete particles to trace 
the phase-space evolution of a continuous density field under the 
influence of gravity. 
They have broad applications in modern 
cosmology that range from structures beyond galaxy clusters
\citep[e.g.][]{pb02, bhb04} to earth-mass dark-matter halos emerging 
as the first objects in the universe \citep*{dms05}. 

Because of their Monte Carlo nature, $N$-body codes have to 
soften the gravity to subdue destructive effects of strong 
two-body scatterings and to increase numerical efficiency \citep{d01}. 
Although softening reduces the variance of the force from discrete 
particles, it also introduces a bias \citep{m96}.
The bias increases with the softening length, while the variance
the opposite. As such, there must exist an optimal softening length
that reaches the best compromise between the bias error and variance
error.

Suitable softening lengths are often searched through convergence 
tests with $N$-body halo simulations 
(\citealt*{nfw96}; \citealt{mgq98}; \citealt{sms98}; \citealt{kkg00}; 
\citealt{fm01}; \citealt{pnj03}). 
It should be emphasized that a proper softening length must be 
matched with other simulation parameters such as the time step.
For example, with a poor combination of the softening length and 
time step, the inner halo profile could become core-like 
\citep*[][hereafter FKM04]{fkm04}.
It is not practical to search for every possible softening 
length and its matching simulation parameters using $N$-body 
simulations. Thus, the resulting softening length may not be 
optimal, and it is not clear what physical error bounds this 
softening length imposes.

\citet{m96} devised a more efficient and objective method,
which requires the optimal softening length to minimize the mean
integrated square error of the force:
\be
{\rm MISE} = \int \rho(\mathbf{r}) \big \langle 
|\mathbf{F}(\mathbf{r}) - \mathbf{F}^{\rm True}(\mathbf{r}) |^2 
\big \rangle {\rm d} \mathbf{r},
\ee
where $\langle \ldots \rangle$ stands for an ensemble average, and
$\rho(\mathbf{r})$ is the continuous density. The true force refers 
to the Newtonian result in the continuous density field.
The MISE is effectively a sum of mass-weighted square bias 
and variance, i.e.
\bea \nonumber
{\rm MISE} &=& \int \rho(\mathbf{r}) 
| \langle \mathbf{F}(\mathbf{r})\rangle -
\mathbf{F}^{\rm True}(\mathbf{r}) |^2 {\rm d} \mathbf{r} + \\ 
\nonumber && \int \rho(\mathbf{r}) 
[\langle \mathbf{F}^2(\mathbf{r})\rangle - 
\langle \mathbf{F}(\mathbf{r})\rangle^2] {\rm d} \mathbf{r}.
\eea

For halo simulations,
the bias error is significant only at the center of the halo, while 
the variance error decreases relatively slowly outward. Hence, by
minimizing the MISE one tends to allow large bias errors in the center
in order to match small variance errors integrated over the whole 
halo, which may not be desirable for halo simulations. To ensure the 
accuracy of the density profile beyond a small radius $r_0$, one may 
require that the bias and variance errors, rather than the 
integrated ones, be both less than a threshold for $r > r_0$.
This is the basis of our approach for optimizing the softening length.

The MISE method has been implemented for halos with $N$-body force 
evaluations at all radii \citep[e.g.][]{m96, afl00, d01}, which demands
much less run time than full $N$-body halo simulations. For a very large 
number of particles, however, a direct $N$-body summation of forces 
could still be time-consuming. To improve the efficiency, we derive 
semi-analytic expressions for the ensemble-average bias and variance 
errors assuming a Poisson sampling of the halo density profile with 
particles. In this work, we use spherically symmetric halos as targets 
for calculating the acceleration error and optimizing the softening 
length. Our method may be generalized for broader applications.

\section{Acceleration Bias} \label{sec:bias}

\begin{figure}
\centering
\includegraphics[width=65mm]{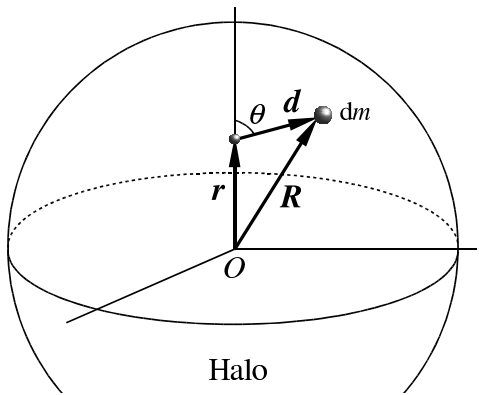}
\caption[f1]{The configuration for calculating the acceleration on
a particle at a distance $r$ from the center of the halo. 
\label{fig:conf}}
\end{figure}

For convenience, we consider the acceleration bias and variance and
express lengths in units of the halo virial radius $r_{\rm v}$. 
We set the mass of the halo within $r_{\rm v}$ to 1, so that the mass 
of each particle is $m_{\rm p} = N^{-1}$, where $N$ is the 
number of particles within $r_{\rm v}$.

The gravitational attraction between two particles can be generalized 
as $F = m_{\rm p}^2 f(r, \epsilon)$,
where we have dropped Newton's constant. For Newtonian gravity
$f(r,\epsilon) = r^{-2}$, and for Plummer softening
$f(r, \epsilon) = (r^2 + \epsilon^2)^{-1}$.
We also utilize the S2 softening \citep{he81}, which treats 
particles as spheres of radius $\epsilon/2$ with density decreasing
linearly from the maximum at the center to zero at 
$\epsilon/2$. 
The S2 softening is often used in particle-particle 
particle-mesh codes \citep[e.g.][]{c91,jf94}.

Suppose that a particle is located at a distance $r$ from the 
center of the halo as illustrated in Fig.~\ref{fig:conf}. 
The mean radial acceleration of the particle is 
\be \label{eq:ar-def}
\langle a_{\rm r}\rangle = \big \langle \int f(d, \epsilon) 
\cos \theta {\rm d} m \big \rangle,
\ee
where ${\rm d}m = m_{\rm p} n(\mathbf{R}) {\rm d}V$ with 
$n(\mathbf{R})$ and ${\rm d}V$ being the particle number density and
volume element at $\mathbf{R}$, respectively. By definition, we have
\be \label{eq:rho}
\rho(R) \simeq m_{\rm p} \langle n(\mathbf{R})
\rangle_{|\mathbf{R}| = R},
\ee
where $\rho(R)$ is the spherically symmetric halo density.
The number of particles 
$n(\mathbf{R}){\rm d}V$ within the volume ${\rm d}V$ follows
approximately the Poisson distribution so that its mean and 
variance are both $N \rho(R) {\rm d}V$. In addition, 
$n(\mathbf{R}){\rm d}V$ at two locations are 
uncorrelated, so that
\bea \label{eq:cov}
m_{\rm p}^2 \langle n(\mathbf{R}) n(\mathbf{R}') \rangle {\rm d}V 
{\rm d}V' &\simeq& \rho(R) \rho(R') {\rm d}V {\rm d}V' + 
\\ \nonumber 
&& \delta^{\rm D}(\mathbf{R}-\mathbf{R}')
N^{-1} \rho(R) {\rm d}V {\rm d}V',
\eea
where $\delta^{\rm D}(\mathbf{R})$ is the Dirac Delta function.

\begin{figure}
\centering
\includegraphics[width=65mm]{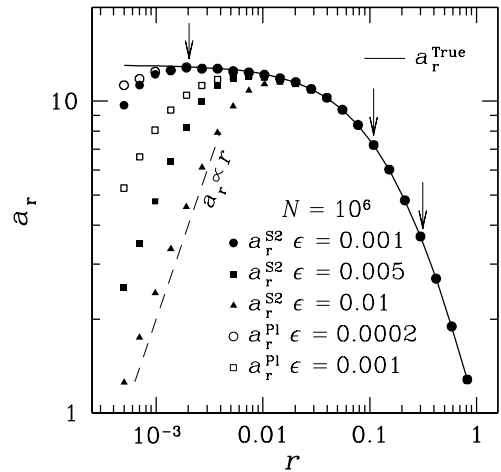}
\caption[f1]{Radial accelerations in an NFW halo of 
concentration $c = 5$. 
The halo is populated with identical particles of mass 
$10^{-6}$ out to twice its virial radius $r_{\rm v}$, and the 
total mass within $r_{\rm v}$ is $1$. The results are obtained 
over 10,000 random realizations of the halo in each case.
The true value $a_{\rm r}^{\rm True}$
(solid line) is calculated for a continuous halo density with 
Newtonian gravity. Solid symbols correspond to the S2 softening
with softening length $\epsilon = 0.001$ (circles),
$\epsilon = 0.005$ (squares), and $\epsilon = 0.01$ (triangles).
Open symbols correspond to Plummer softening with 
$\epsilon = 0.0002$ (circles) and $\epsilon = 0.001$ (squares).
The $a_{\rm r}\propto r$ behavior of the radial acceleration in a 
constant density core is drawn in a dashed line for comparison. 
The downward arrows mark the radii where the mean interparticle 
distance is equal to the softening length in the S2 cases.
\label{fig:bias}}
\end{figure}

Combining equations (\ref{eq:ar-def}) and (\ref{eq:rho}) one gets
\be \label{eq:ar}
\langle a_{\rm r}\rangle = \pi \int_{0}^{\pi}\! \sin 2\theta 
{\rm d} \theta
\int_0^{d_{\theta}} \! f(d, \epsilon) \rho(R) d^2 {\rm d} d,
\ee
where $R^2 = d^2 + r^2 + 2 d r \cos \theta$, and the integral limit 
$d_{\theta}$ is solved from
$d_{\theta}^2 + 2 d_{\theta} r \cos \theta + r^2 = R_{\rm max}^2$. 
For Newtonian gravity $R_{\rm max} = r$, while for softened gravity 
$R_{\rm max}$ is $r$ plus the range that softening is in effect. 
The bias is then
\be \label{eq:bias}
b_{\rm a} = \langle a_{\rm r} \rangle - a_{\rm r}^{\rm True}.
\ee

\begin{figure*}
\centering
\includegraphics[width=120mm]{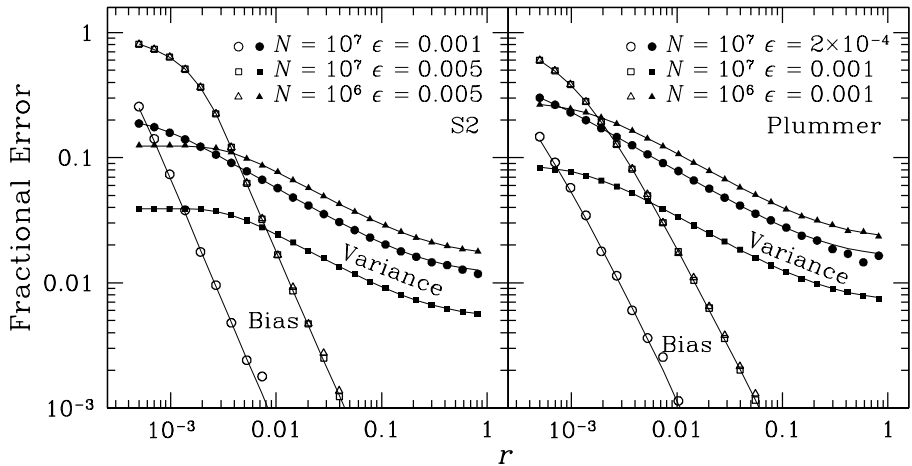}
\caption[f2]{Fractional acceleration bias error
($|b_{\rm a}|/a_{\rm r}^{\rm True}$, open symbols) and variance 
error ($\sigma_{\rm a}/a_{\rm r}^{\rm True}$, solid symbols) as 
functions of the distance from the center of the halo. 
Solid lines that trace the symbols are the corresponding semi-analytic 
results using equation (\ref{eq:bias}) for the bias error and
equation (\ref{eq:va}) for the variance error. 
When the interparticle distance is much larger than the softening
length, one cannot sample the long tail error distribution 
well enough, even with 10,000 realizations, to measure the variance
accurately. This leads to small discrepancies in the variance 
for small $\epsilon$ at $r\sim 1$, which can be reduced 
with more realizations. The bias error does 
not depend on the number of particles $N$ within the virial radius, 
and the variance error scales as $N^{-1/2}$ for the same softening.
\label{fig:bvr}}
\end{figure*}

Fig.~\ref{fig:bias} shows a few examples of 
$\langle a_{\rm r} \rangle$ under 
softened gravity. The accelerations are evaluated with direct
$N$-body force summations over 10,000 random realizations of a 
Navarro-Frenk-White \citep*[NFW,][hereafter NFW97]{nfw97} halo, 
which has a concentration number $c = 5$. As expected, 
the acceleration bias is significant only at 
$r$ less than a few $\epsilon$ and is small at larger radii 
even where particles can be, on average, closer than the 
softening length in the S2 case.

With softening lengths of a few percent of the virial 
radius\footnote{The softening splines in NFW97 and N04 differ from 
S2, but we do not expect such differences to alter the result 
qualitatively.} 
\citep[e.g.~NFW97;][hereafter N04, see Table \ref{tab:err}]{nhp04},
the radial acceleration at $r < 0.01 r_{\rm v}$ behaves as if 
there was a constant density core, even though the underlying 
density has a central cusp with a logarithmic slope of $-1$. 
Plummer softening performs worse than the S2 
softening at the same quoted softening length, because it has a
broader softening kernel than S2 does. 

For better comparisons, we present fractional bias errors 
$|b_{\rm a}|/a_{\rm r}^{\rm True}$ in Fig.~\ref{fig:bvr}, which
includes $N = 10^7$ results from 10,000 realizations of the same 
halo. There is a good agreement between direct $N$-body force 
summations (open symbols) and the semi-analytic results of equation 
(\ref{eq:bias}) (solid lines). Note that the bias error does not
depend on the number of particles within the virial radius. 

\section{Acceleration Variance} \label{sec:var}

At a fixed position, the acceleration 
fluctuates from one realization to another because of the 
discrete sampling of the halo by particles. These 
fluctuations imposes a sample variance error on 
particles' acceleration, and they depend on both the number
of particles and softening. With the help of equation 
(\ref{eq:cov}) we find the acceleration variance
\be \label{eq:va}
\sigma_{\rm a}^2 = \langle \mathbf{a}^2 \rangle - \langle 
\mathbf{a} \rangle^2 = \frac{2\pi}{N} \int_{0}^{\pi} \! 
\sin \theta {\rm d} \theta
\int_0^{d_{\theta}} \! f^2(d, \epsilon) \rho(R) d^2 {\rm d} d,
\ee
where 
$d_{\theta}^2 + 2 d_{\theta} r \cos \theta + r^2 = R_{\rm cut}^2$,
and $R_{\rm cut}$ formally extends to infinity. Since the result
within the virial radius converges very quickly for 
$R_{\rm cut} \gtrsim$ a few $r_{\rm v}$, we set
$R_{\rm cut} = 2 r_{\rm v}$ to be consistent with $N$-body 
calculations. 

We calculate fractional variance errors 
$\sigma_{\rm a}/a_{\rm r}^{\rm True}$ for the same set of 
configurations as for fractional bias errors in \S \ref{sec:bias}. 
The variance errors of direct $N$-body force summations are plotted 
in Fig.~\ref{fig:bvr} with solid symbols. One sees that the $N$-body
results are very well traced by equation (\ref{eq:va}) in solid 
lines. For a fixed $N$, a smaller softening length results in 
larger variance errors, while, for a fixed softening length, the
variance error scales as $N^{-1/2}$. Plummer softening has 
smaller variance errors than the S2 softening with the same 
$\epsilon$ and $N$ because of its broader softening kernel.


To access the effect of the halo profile, we calculate acceleration
errors using different concentration numbers and using
\citet[][hereafter M99]{mqg99} profile, which has a stronger cusp
of logarithmic slope $-1.5$.
The bias error does not change much with the profile. 
On the other hand, the variance error is reduced by a factor of 2 to 
3 by boosting the concentration from $c = 5$ to $c = 20$. 
The variance error of M99 halos is a factor of 1.6 smaller than 
(roughly the same as) its corresponding NFW halos at $r = 0.01$ 
($r \sim 1$), when the concentration number is adjusted to
$c({\rm M99}) = [c({\rm NFW})/1.7]^{0.9}$ \citep{ps00}. 

\begin{deluxetable}{l c c c c}
\tablewidth{0pt}
\tablecaption{Error Estimates for Selected Simulations
\label{tab:err}}
\tablehead{\colhead{ } & 
\colhead{\phn $\tilde{\epsilon}$} &
\colhead{$N$} & 
\colhead{\phn $|b_{\rm a}|/a_{\rm r}^{\rm True}$} & 
\colhead{$\sigma_{\rm a}/a_{\rm r}^{\rm True}$} \\
\colhead{ } & 
\colhead{\phd $(\times 10^{-3})$} &
\colhead{} & \colhead{\phn $(\%)$\phd} & 
\colhead{$(\%)$\phd}}
\startdata
NFW97 & \phd 20 & $10^{4}$ & \phn \phn \phd 30 \phn & 30 \\ 
M99   & \phn \phd 1 & $10^{6}$ & \phn 0.07 \phn & \phn 8 \\ 
N04   & \phd 10 & $10^{6}$ & \phn\phn \phd \phn 7 \phn & \phn 5 \\ 
FKM04 & 0.3  & $10^{7}$ & \phn \phn 0.2 \phn & \phn \phn 6
\enddata
\tablecomments{The acceleration errors are estimated at $1\%$ of
the virial radius. For spline softening (NFW97, M99, and 
N04), $\tilde{\epsilon}$ equals the scale beyond which the gravity
is Newtonian. Whereas for Plummer softening (FKM04), 
$\tilde{\epsilon}$ is their reported softening length.
The number of particles $N$ within $r_{\rm v}$ is approximate.
}
\end{deluxetable}

\begin{figure*}
\centering
\includegraphics[width=120mm]{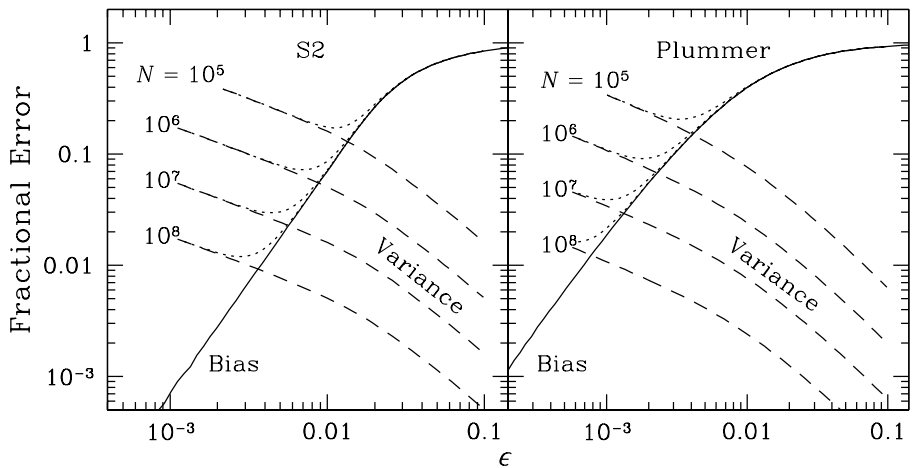}
\caption[f3]{Fractional acceleration bias error (solid lines), 
variance error (dashed lines), and total error (dotted lines) at 
$r = 0.01r_{\rm v}$ as functions of the softening length.
The optimal softening length is set by the minimum of
the total acceleration error.
\label{fig:bve}}
\end{figure*}

\section{Optimal Softening} \label{sec:opt}
As seen in Fig.~\ref{fig:bvr}, the bias error increases much faster
toward the center of the halo than the variance error. Minimizing
the MISE \citep{m96} may not be optimal for studying inner
halo properties, because the MISE method prefers a relatively 
large softening length so that small variance errors spread over the 
entire halo are matched by large bias errors confined in the center.

We propose to optimize the softening length by minimizing 
the mean acceleration error 
$\mathcal{E}_{\rm a} = \sqrt{b_{\rm a}^2 + \sigma_{\rm a}^2}$
at a small radius $r_0$. In this way, one
ensures that $\mathcal{E}_{\rm a}$ never exceeds the optimal 
value at larger radii. Conversely, one could set an error budget
at $r_0$ and determine the number of particles needed 
for a particular softening. 

Fig.~\ref{fig:bve} illustrates the mean acceleration error at $1\%$
of the halo virial radius (dotted lines) as a function of the 
softening length for different number of particles. 
For an NFW ($c=5$) halo with S2 softening, we obtain, 
from the minimum of each error curve,
$\epsilon_{\rm opt} = 0.11 N^{-0.20}$ and 
$\mathcal{E}_{\rm a} = 14 N^{-0.38}$. For the same halo but 
with Plummer softening $\epsilon_{\rm opt} = 0.065 N^{-0.26}$ and 
$\mathcal{E}_{\rm a} = 13 N^{-0.36}$. The optimal softening lengths
and minimum acceleration errors of corresponding M99 halos have 
nearly the same $N$-dependence as those of NFW halos but with 
15--30\% smaller prefactors, because they have smaller variance 
errors at the same soften length.

%

The $N$-dependence of our results are consistent with those of 
MISE results in \citet{afl00} and \citet{d01}, despite that our 
halos differ from theirs and that we prefer smaller softening 
lengths. For a closer comparison, we calculate the optimal 
softening length for a \citet{h90} sphere 
of $10^5$ particles with Plummer softening. 
We find $\epsilon_{\rm opt} = 0.0049$ 
and $\mathcal{E}_{\rm a} = 30\%$ with $r_0 = 0.01$.
Whereas \citet{d01} obtained $\epsilon_{\rm opt} = 0.016$ and 
a mass-weighted average error of $6.7\%$, which lead to
$\mathcal{E}_{\rm a} = 55\%$ at $r = 0.01$. With our 
$\epsilon_{\rm opt}$ the average error is $7.2\%$. This 
shows that our strategy is to trade slightly larger  
(yet tolerable) errors at large radii with smaller errors near 
the center (though $10^5$ particles do not seem 
sufficient for studying inner halo properties no matter
what softening length is used).

\section{Discussion and Conclusions} \label{sec:dis}
The optimal softening lengths in this work
are not truly optimal in the sense that they slightly 
depend on the assumed halo profile and that, in turn, the simulated 
halo profile could be affected by the softening length. 
Nevertheless, our strategy provides guidance for choosing
the softening length and for evaluating acceleration errors of
simulated halos. For example, we list in Table~\ref{tab:err}
error estimates for four sets of halo 
simulations ranging from one of the first claims of the universal 
profile (NFW97, $N < 10^4$) to the latest 
investigation (FKM04, $N > 10^7$). The errors 
are typically $\gtrsim 6\%$ at $r = 0.01r_{\rm v}$. Since the 
mass within $0.01 r_{\rm v}$ is a few thousandths of the halo 
virial mass, even a few percent acceleration errors may contribute 
significantly to the uncertainties in the inner slope of the halo.
In fact, N04 achieve convergent results only at 
$r \gtrsim 0.01 r_{\rm v}$. Hence, it may not be reliable to
extrapolate to ever smaller radii and infer a central cusp.


\citet{pnj03} propose that the optimal softening length should 
satisfy the condition that the maximum stochastic acceleration 
($\sim 1/N\epsilon^2_{\rm opt}$) be several factors smaller than
the mean field acceleration at $r_{\rm v}$ ($\sim 1/r^2_{\rm v}$). 
Roughly speaking, the acceleration variance arises from two sources: 
the stochastic acceleration as defined by \citet{pnj03} and  
Poisson fluctuations of particles in the halo. 
Since the latter increases toward the center, setting a 
small stochastic acceleration at $r_{\rm v}$ does not 
always guarantee small errors near the center. 

Our criterion for the optimal softening length imposes a strict
upper limit on the mean acceleration error at $r > r_0$. 
Direct error bounds on halo properties may be more useful for 
interpreting simulation results, but, to identify the most
accurate simulation upon which error estimates of other simulations 
can be based, one needs a gauge like the mean acceleration error.

To optimize the softening length for general density fields or halos 
with sub-structures or asymmetries, one must derive equations 
(\ref{eq:ar}) and (\ref{eq:va}) without assuming spherical symmetry 
for the density. Moreover, one should generally minimize the 
acceleration error in high density regions where the error tends to 
be large.

From NFW97 to N04, the acceleration error at $0.01 r_{\rm v}$ is 
reduced from $42\%$ to $8.6\%$, and one starts to see a continuously
changing (logarithmic) central density slope instead of a constant 
slope of $-1$ in an NFW halo. Hence, it will be interesting to see
how the results evolve when the error is further reduced to well
below $1\%$ with $N \gg 10^8$. 
A recent investigation on two-body relaxation also suggests
that $\gg 10^8$ particles are required to faithfully model the very 
inner regions of halos \citep{e05}.

\acknowledgements
HZ was supported by the National Science Foundation under 
Grant No. 0307961 and 0441072 and NASA under grant No. NAG5-11098.


\end{document}